\def\year{2022}\relax
\documentclass[letterpaper]{article} 
\usepackage{aaai22}  
\usepackage{times}  
\usepackage{helvet}  
\usepackage{courier}  
\usepackage[hyphens]{url}  
\usepackage{graphicx} 
\usepackage{natbib}  
\usepackage{caption} 
\DeclareCaptionStyle{ruled}{labelfont=normalfont,labelsep=colon,strut=off} 
\usepackage{multirow}
\usepackage{booktabs}       
\usepackage{amsmath,graphicx}
\usepackage{amssymb}
\usepackage{algorithm}
\usepackage{hyperref}
\usepackage{algorithmic}
\usepackage[table, svgnames, dvipsnames]{xcolor}

\newcommand{\so}[1]{} 
\definecolor{mygreen}{HTML}{008000}

\usepackage{newfloat}
\usepackage{listings}
\floatstyle{ruled}
\newfloat{listing}{tb}{lst}{}
\floatname{listing}{Listing}

\title{Self-Supervised Representation Learning for Speech Using Visual Grounding and Masked Language Modeling}
\author{
    Puyuan Peng, David Harwath
}
\affiliations{
    Department of Computer Science, The University of Texas at Austin\\

    Austin, Texas, 78712, USA \\
    pyp@utexas.edu, harwath@utexas.edu\\

}

\begin{document}

\maketitle

\begin{abstract}
In this paper, we describe our submissions to the ZeroSpeech 2021 Challenge and SUPERB benchmark. Our submissions are based on the recently proposed FaST-VGS model, which is a Transformer-based model that learns to associate raw speech waveforms with semantically related images, all without the use of any transcriptions of the speech. Additionally, we introduce a novel extension of this model, FaST-VGS+\footnote{code and model weights are available at \href{https://github.com/jasonppy/FaST-VGS-Family}{https://github.com/jasonppy/FaST-VGS-Family}}, which is learned in a multi-task fashion with a masked language modeling objective in addition to the visual grounding objective. On ZeroSpeech 2021, we show that our models perform competitively on the ABX task, outperform all other concurrent submissions on the Syntactic and Semantic tasks, and nearly match the best system on the Lexical task. On the SUPERB benchmark, we show that our models also achieve strong performance, in some cases even outperforming the popular wav2vec2.0 model.
\end{abstract}

\section{Introduction}
With the popularization of self-supervised learning (SSL), the speech community has made significant progress towards reducing the amount of labelled data needed to reach strong levels of performance in a variety of tasks. To better measure the progress within this research area, two benchmarks, the ZeroSpeech 2021 challenge~\cite{zs21,alishahi2021zr2021vg} and the SUPERB benchmark~\cite{yang2021superb}, were recently proposed. 

The two benchmarks both aim at examining the properties of the representations learned in SSL models, but from different angles. ZeroSpeech2021 aims at probing the language modeling capabilities of acoustic models in the zero-resource setting. To achieve this, its tasks measure how well the representations from a pretrained model capture an utterance's phonetic content, lexical correctness, syntactic correctness, and semantics. On the other hand, SUPERB provides a standardized system for benchmarking SSL models' performance on a range of speech tasks including speaker identification and diarization, automatic speech recognition, emotion recognition, and more. SUPERB evaluates models by learning task-specialized lightweight prediction heads on top of the frozen pretrained SSL model.

While there have been encouraging progress on the two benchmarks, most of the SSL models are trained using only speech data. However, humans learn language within a rich multimodal context, where vision is an important component. In this work, we explore the benefit of incorporating visual context for self-supervised speech representation learning. In particular, we study a recently proposed model FaST-VGS~\cite{peng2021fastslow} which is trained via a cross-modal contrastive loss to match visual images with the speech that describes them. In this paper, we also introduce a novel extension of FaST-VGS which we call FaST-VGS+, that augments FaST-VGS with a wav2vec2.0-type~\cite{Baevski2020wav2vec2A} masked language modeling loss applied to the speech audio. We test the two models on the a speech-image retrieval benchmark, ZeroSpeech2021, and SUPERB. 

\section{Related Work}
\textbf{SSL for speech}. SSL for speech has recently become extremely popular. Most current techniques are based upon predicting some part of the speech signal conditioned on some other part of the signal, using standard prediction losses such as L2 distance or cross-entropy~\cite{Chung2019AnUA,chung2020vectorquantized,Liu2020MockingjayUS,tera,Ling2020DeCoAR2D,hsu2021hubert,chen2021wavlm}, using a contrastive objective~\cite{infonce,schneider2019wav2vec,baevski2020vqwav2vec,Baevski2020wav2vec2A}, or multi-task learning~\cite{pascual2019learning,ravanelli2020multitask}. It has been shown that SSL models can achieve state-of-the-art or very competitive results on automatic speech recognition~\cite{Baevski2020wav2vec2A}, phoneme
classification~\cite{Chung2019AnUA}, speaker identification~\cite{fan2021exploring}, emotion recognition~\cite{pascual2019learning},
speech translation~\cite{Chung2019AnUA}, spoken language understanding~\cite{lai2020semisupervised},
voice conversion~\cite{lin2021fragmentvc} and text-to-speech translation~\cite{problemagnostic}. The SUPERB benchmark is the first effort that attempts standardize the measurement of progress on SSL for speech. In addition to the initial benchmark systems curated by the SUPERB organizers, there have been several recent, notable additions to the leaderboard. DistilHuBERT~\cite{chang2021distilhubert} is a multitask knowledge distillation model built upon HuBERT~\cite{hsu2021hubert} that drastically reduces the number of model parameters and inference time, while retains most of the performance on SUPERB. WavLM~\cite{chen2021wavlm} is also built upon HuBERT with two additional techniques, namely gated relative position bias~\cite{chi2021xlme} and utterance mixing. Trained on 94k hours of audio data, WavLM is able to outperform all prior work on all ten SUPERB tasks. Instead of dramatically increasing the magnitude of audio training data, we tackled the problem from a different perspective - adding multimodal information using a visual grounding objective - and we show that on several tasks, our models can outperform models trained on much larger amounts of speech data. 

\textbf{Visually grounded speech (VGS) processing}. Visually-grounded speech processing has recently attracted an increasing amount of attention~\cite{chru2021visually}. Starting from the work of~\citet{Synnaeve14learningwords,Harwath2015DeepMS}, researchers have studied the ability of models to learn to recognize the structure of spoken language, such as words and sub-word units, by training the models to associate speech waveforms with contextually relevant visual inputs. These works have looked at a variety of tasks, such as speech-image retrieval~\cite{Harwath2016UnsupervisedLO,Chrupaa2019SymbolicIB,Ilharco2019LargescaleRL,Mortazavi2020SpeechImageSA,Sanabria2021TalkDW}, automatic speech recognition~\cite{Sun16LookListenDecode,Palaskar18GroundedASR,Hsu2019TransferLF}, word detection and localization~\cite{Kamper2017VisuallyGL,Harwath2017LearningWU,Merkx2019LanguageLU,Wang2020ADH,Olaleye2021AttentionBasedKL}, hierarchical linguistic unit analysis~\cite{Chrupaa2017RepresentationsOL,Harwath2020LearningHD}, cross-modality alignment~\cite{Harwath2019JointlyDV,Wang2021AlignOA,khorrami2021evaluation}, speech segmentation~\cite{Harwath2019TowardsVG}, speech generation~\cite{Hsu2020TextFreeIS}, and learning multilingual speech representations~\cite{Harwath2018VisionAA,Kamper2018VisuallyGC,Havard2020CatplayinginthesnowIO,Ohishi2020TrilingualSE}. In this paper, we study the recently proposed FaST-VGS~\cite{peng2021fastslow} speech-image retrieval model, and and propose a novel extention of the model that incorporates a wav2vec2.0-style~\cite{Baevski2020wav2vec2A} masked language modeling objective in a multi-task learning framework.

\textbf{The ZeroSpeech Challenge}. Since 2015, the ZeroSpeech challenge has acted as a benchmark to assess the ability of unsupervised speech models to learn the structure of spoken language in the so-called ``zero-resource'' setting where no conventional annotations (i.e. text transcriptions) are available. Previous challenge tasks have included speech segmentation, spoken unit discovery, and speech synthesis. The ZeroSpeech 2021 challenge adds several new tasks centered on performing language modeling directly on the speech signal. The challenge contains two tracks, the non-visually-grounded track and the visually-grounded (VG) track. The non-VG track was introduced earlier, and submissions to this track studied the effect of filtering out speaker information~\cite{Niekerk21}, denoising CPC~\cite{infonce} representations by using methods from information retrieval~\cite{Chorowski2021InformationRF}, and combining CPC with deep clustering~\cite{Maekaku2021SpeechRL}. These ideas offered improved results on the phonetic, lexical and syntactic tasks. However, the semantic task has thusfar seen less progress. The VG track was more recently proposed investigate how incorporating visual context could potentially help performance on the spoken language modeling tasks. The challenge introduced several baseline  VG models~\cite{Higy2020TextualSF,Chrupaa2019SymbolicIB}, which achieve improvements over the current non-VG models on the semantic tasks, but fall somewhat short on other three tasks. In this work, we show that our model significantly outperforms these baselines and existing submissions on the semantic tasks, while also performing at or near the top on the other three tasks.

\section{Models}
In this section, we describe two models, namely FaST-VGS and FaST-VGS+. FaST-VGS is a transformer-based VGS model proposed by~\cite{peng2021fastslow}, which is trained using a contrastive speech-image retrieval loss~\citep{Ilharco2019LargescaleRL}. FaST-VGS+ augments FaST-VGS with a wav2vec2.0-style masked language modeling loss~\citep{Baevski2020wav2vec2A}. As we show in our experimental results, FaST-VGS+ performs as well as FaST-VGS on speech-image retrieval, and outperforms FaST-VGS on all four tasks in the ZeroSpeech2021 Challenge, as well as eight out of ten tasks on the SUPERB Benchmark.

\begin{figure*}
     \centering
     \includegraphics[width=0.9\textwidth]{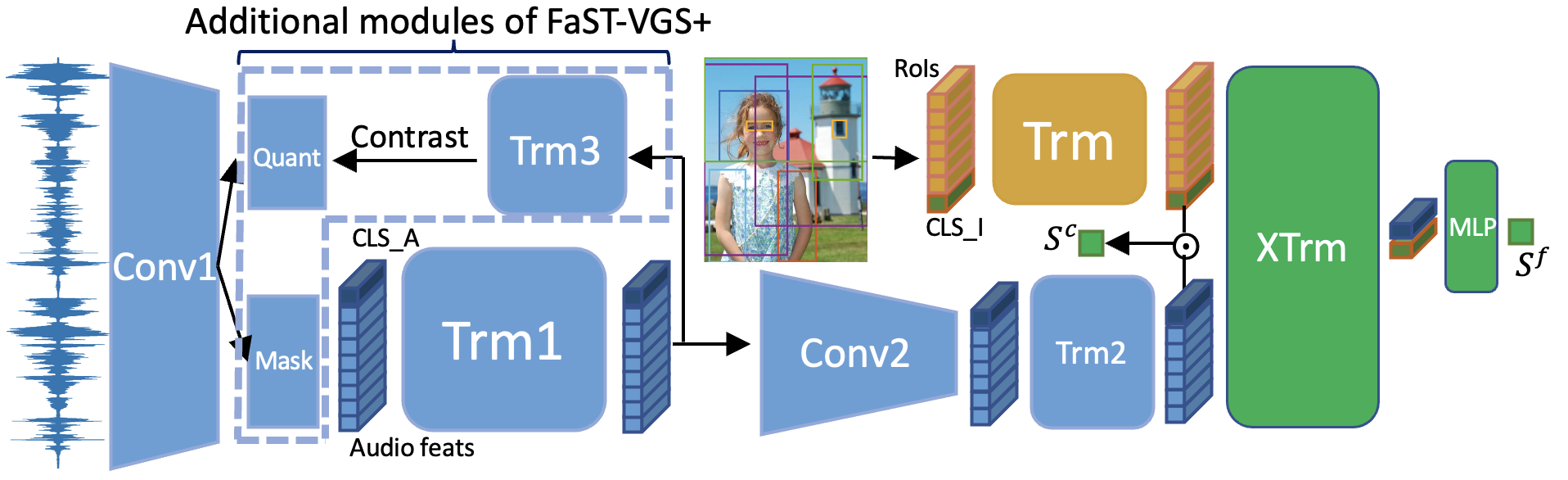}
     \caption{FaST-VGS and FaST-VGS+. The ``Mask,'' ``Quant,'' and ``Trm3'' modules inside the blue dashed lines differentiate the FaST-VGS model from FaST-VGS+. The object detector module that produces the set of RoIs from an input image is omitted in the figure.}
     \label{fig:archi}
 \end{figure*}

\textbf{FaST-VGS}. The input to FaST-VGS is a visual image $I$ and an acoustic speech waveform $A$ that describes the contents of image $I$. The model (Fig.~\ref{fig:archi}) contains an audio encoder, an image encoder, and a cross-modal encoder. \texttt{ClS\_I} and \texttt{ClS\_A} tokens are concatenated with features of $I$ and $A$ at an early stage during forward propagation, and are used to produce similarity scores between image and audio. The similarity scores should be high when the audio describes the contents of the image, and low otherwise. In particular, FaST-VGS produces two similarity scores for every input pair --- the coarse similarity $S^c$, which is calculated by taking the dot product of the output \texttt{ClS\_I} and \texttt{ClS\_A} of the two unimodal encoders; and the fine similarity $S^f$, which is calculated by passing the output \texttt{ClS\_I} and \texttt{ClS\_A} of the cross-modal encoder to an MLP. As demonstrated by \citet{peng2021fastslow}, retrieval based on the fine similarity $S^f$ is more accurate than retrieval based on the coarse similarity $s^c$, but can be much slower, especially when the search database is large. To combine the advantages of fast retrieval using $S^c$ and accurate retrieval using $S^f$, \citet{peng2021fastslow} proposes the coarse-to-fine retrieval technique (CTF). CTF uses a two-pass strategy by using $S^c$ to retrieve a set of $K^c$ target items from the search database, and only passes the features of these items and the features of the query to the cross-modal encoder. The fine scores between these pairs are then used for the final retrieval. \citet{peng2021fastslow} shows that CTF retrieval can achieve the same accuracy as fine retrieval, while approaching the retrieval speed of coarse retrieval.

The training objective of the FaST-VGS is a cross-modal constrastive loss that encourages the similarity scores of matched speech and image pairs to be high and that of unmatched pairs to be low. To this end, a masked and marginalized version of the InfoNCE~\cite{infonce} loss introduced by~\citet{Ilharco2019LargescaleRL} is used on both $S^c$ and $S^f$ to compute the coarse matching loss and the fine matching loss. Mathematically, given a batch of $B$ pairs of images and audio captions, the similarity score $S^*_{i,j}$ ($*\in\{c,f\}$) between the $i^{th}$ audio and the $j^{th}$ image is calculated, and the loss is defined as $$\mathcal{L}^* = \mathcal{L}^*_{A\rightarrow{I}} + \mathcal{L}^*_{I\rightarrow{A}},$$ where  
\begin{eqnarray}
    \mathcal{L}^*_{A\rightarrow{I}} = -\frac{1}{B}\sum_{i=1}^{B}\log\frac{e^{S^*_{i,i}-\delta}}{e^{S^*_{i,i}-\delta} + \sum_{j=1}^{B}M_{i,j}e^{S^*_{i,j}}} \\
    \mathcal{L}^*_{I\rightarrow{A}} = -\frac{1}{B}\sum_{i=1}^{B}\log\frac{e^{S^*_{i,i}-\delta}}{e^{S^*_{i,i}-\delta} + \sum_{j=1}^{B}M_{j,i}e^{S^*_{j,i}}}
\end{eqnarray}
We assume that the $i^{th}$ audio caption and the $j^{th}$ image are semantically matched. Since the SpokenCOCO~\citep{Hsu2020TextFreeIS} dataset contains 5 audio captions for each image, we need to prevent these pairs being treated as negative examples. To this end, in the denominator of the loss terms, we use masking variables $M_{i,j}$ which are $0$ when audio caption $i$ matches image $j$, and $1$ otherwise. The FaST-VGS is trained to optimize both coarse and fine matching loss simultaneously:
$$\mathcal{L}_{\text{FaST-VGS}} = \lambda^c\mathcal{L}^{c} + \lambda^f\mathcal{L}^{f}.$$

\citet{peng2021fastslow} found that putting a larger weight on the fine matching loss during training is helpful for coarse retrieval, fine retrieval, and CTF retrieval.

\textbf{FaST-VGS+} (Fig.~\ref{fig:archi}). We now extend FaST-VGS by incorporating a wav2vec2.0-style~\citet{Baevski2020wav2vec2A} masked language modeling loss on the speech signal. To do so, at the end of Conv1, a masking module is used to randomly mask $p\%$ of the tokens, which are then concatenated with \texttt{CLS\_A} and fed into the first Transformer block Trm1 for contextualization. The output of the Trm1 block is fed as input to the Conv2 block, and also fed into another Transformer block Trm3, whose output is used to predict the quantized and masked Conv1 features. Mathematically, assuming time step $t$ of the audio features extracted by Conv1 is masked, we denote the corresponding output of Trm3 as $c_t$ and the quantized Conv1 feature as $q_t$. The wav2vec2.0 contrastive loss in this case is defined as 
\begin{equation}
    \mathcal{L}^w = -\log \frac{\exp(sim(c_t, q_t)/\kappa)}{\sum_{\tilde{q}_t \in \mathbf{Q}_t} \exp(sim(c_t, \tilde{q}_t)/\kappa)}
\end{equation}
where $\mathbf{Q}_t$ contains both $q_t$ and $K$ distractor examples sampled from elsewhere in the batch. $sim(a,b) = a^Tb/\|a\|\|b\|$ is the cosine similarity.

Following~\cite{Baevski2020wav2vec2A}, to encourage equal use of the codebook entries, we also make use of the codebook diversity loss:
\begin{equation}
    \mathcal{L}^d = \frac{1}{GV} \sum_{g=1}^{G} - H(\bar{p}_{g}) = \frac{1}{GV} \sum_{g=1}^{G} \sum_{v=1}^{V} \bar{p}_{g,v} \log{\bar{p}_{g,v}}
\end{equation}
where $\bar{p}_{g,v}$ is the probability of an input vector being assigned to entry $v$ in codebook $g$, and $\bar{p}_{g}$ is the code distribution of codebook $g$. $G$ is the number of codebooks and $V$ is the number of entries in each codebooks.  The overall loss of FaST-VGS+ is a combination of the four loss introduced above:
$$\mathcal{L}_{\text{FaST-VGS+}} = \lambda^c\mathcal{L}^{c} + \lambda^f\mathcal{L}^{f} + \lambda^w\mathcal{L}^{w} + \lambda^d\mathcal{L}^{d}$$

In the following section, we examine the properties of the representations learned in FaST-VGS and FaST-VGS+ by testing them on three benchmarks: speech-image retrieval on SpokenCOCO, the ZeroSpeech2021 challenge, and the SUPERB benchmark.

\section{Experimental Setup}\label{sec:setup}
\subsection{Implementation Details}
Conv1, Trm1, and Trm3\footnote{Trm3 is only used in FaST-VGS+} are initialized as the convnet feature extractor, first 8 layers of the Transformer, last 4 layers of the Transformer of the wav2vec2.0 Base~\citep{Baevski2020wav2vec2A} model. We also initialize the Gumbel-softmax layer from the wav2vec2.0 Base model, and use the same masking hyperparameters as the wav2vec2.0 model\footnote{Since we didn't perform dynamic batching to crop every example in the batch to the smallest length, we fixed the batch size and introduced padding. We found that the masking function used in \texttt{fairseq} will crop the number of mask to the smallest one, which in our case will lead to batch dependent number of masks and imbalanced masking (i.e. more portion being masked for short utterances and less portion being masked for long utterances). However, we found that this ``bug'' actually led to similar performance in most of the tasks that we studied except for ASR in SUPERB, where it increased WER by about 1\% absolute}. Conv1 is frozen during training. Conv2 has the same architecture as ResDAVEnet~\cite{Hsu2019TransferLF} except for the dimension of the first layer being changed to match Trm1's hidden dimension of 768. Trm2 is a one layer standard Transformer. We use a frozen Faster-RCNN \citep{Ren2015FasterRT} pretrained on Visual Genome \citep{krishnavisualgenome} to extract the top 36 RoIs for each input image ~\cite{Anderson2018BottomUpAT,Tan2019LXMERTLC}. The RoIs and their bounding box positions are fed to a 6-layer Transformer (Trm), which is randomly initialized. The cross-modal encoder has two blocks, with each block containing a cross-attention layer, a self-attention layer, and finally a feedforward layer. The final MLP that computes $S^f$ contains three fully connected layers with dimensions $768$, $1536$, and $1$, and GELU~\citep{hendrycks2016gelu} activation functions. The dimension of all transformers is $768$. For $\lambda^c$ and $\lambda^f$, we follow~\cite{peng2021fastslow} and set them to be $0.1$ and $1$ in all experiments. We also set $\lambda^w = 1$ and $\lambda^d = 0.1$.

\subsection{Datasets and Tasks}
For training, we make use of both the SpokenCOCO \citep{Hsu2020TextFreeIS} and LibriSpeech~\cite{Panayotov2015LibrispeechAA} datasets. SpokenCOCO contains 123k images each with 5 spoken captions produced by humans speaking aloud the text captions from MSCOCO \citep{Lin2014MicrosoftCC}, resulting in a total of $943$ hours of speech audio. LibriSpeech contains $960$ hours of read English speech from public-domain audiobooks. Note that only SpokenCOCO can be used for calculating the visual grounding losses $\mathcal{L}^c$ and $\mathcal{L}^f$, while speech audio from both SpokenCOCO and Libri be used inpeech can be used in the calculation of the losses associated with masked language modeling, $\mathcal{L}^w$ and $\mathcal{L}^d$. Since we will study the effect of using only SpokenCOCO or using both SpokenCOCO and LibriSpeech in our model training, in all of our results we make the training datasets used explicit. Note that since our model loads the weights of wav2vec2.0 base, which is pretrained on LibriSpeech, when we indicate the training data is SpokenCOCO only, it means only SpokenCOCO is used to train the full model.

The ZeroSpeech2021 Challenge aims at evaluating language modeling capabilities of unsupervised speech models in the zero resource setting, where no text transcriptions or other linguistic annotations are provided. For the phonetic ABX task, we use dynamic time warping with frame-level cosine distances. For the semantic task, we use temporal pooling (either mean or max) to get the vector representation of an utterance. For the lexical and syntactic tasks, we follow \citet{alishahi2021zr2021vg} and run the KMeans clustering algorithm with $500$ clusters on the LibriSpeech \texttt{train-clean-100} split. The KMeans model is then used to quantize all 960 hours of the LibriSpeech training set. A RoBERTa model \citep{liu2019roberta} is trained on the quantized LibriSpeech training set with a standard masked language modeling loss \citep{Devlin2019BERTPO}, with mask span size sampled from $\mathcal{N}(5,5)$ and the total masking coverage
is roughly half of the input tokens (spans may overlap). The learning rate is warmed up to a peak value of $2e-5$. The model is trained for at most 250k steps with maximally $4096$ tokens each step. The evaluation of lexical and syntactic tasks also follows \citet{zs21,alishahi2021zr2021vg} which uses a product of multiple spans' probability as the pseudo-probability of the likelihood of a word or sentence. We use the same hyperparameters as \citet{zs21,alishahi2021zr2021vg} (see sec 4.2 in \citet{alishahi2021zr2021vg} for details).

For the phonetic and semantic tasks, since no training is required, we examine the performance of each layer from FaST-VGS and FaST-VGS+. For the lexical and syntactic tasks, due to our limited computational budget, we experiment with layers 6,7, and 8 of the Trm1 module for both FaST-VGS and FaST-VGS+.

Because some components of FaST-VGS and FaST-VGS+ are initialized from wav2vec 2.0, in our experiments we also compare against the wav2vec2.0 model. We also investigate the impact of different datasets. In addition to wav2vec2.0 trained on LibriSpeech, FaST-VGS trained on SpokenCOCO, and FaST-VGS+ trained on SpokenCOCO and LibriSpeech, we also study the performance of wav2vec2.0 trained on both librispeech and the audio captions of SpokenCOCO, and finally FaST-VGS+ trained only on SpokenCOCO. As we show in our results, the use of visual grounding has a significant impact on the model's performance on the ZeroSpeech tasks.

For the SUPERB benchmark, our results are preliminary, as we did not perform any hyperparameter search except for trying a different learning rate and changing the number of training steps for the phoneme recognition task. Otherwise, all of our experiments use the default hyperparameters. We did find that using a lower learning rate and a larger number of training steps improved the performance on the phoneme recognition task. We also found that changing the weight on $\mathcal{L}^w$ from $1$ to $10$ further improved the phone error rate from $7.76$ to $6.68$, but at the expense of other tasks such as intent classification (IC) where the accuracy decreases from $98.37$ to $97.76$. The s3prl toolkit used by SUPERB automatically optimizes for a weighted summation of features from different layers as the input to downstream models; for this purpose, we provide the output of Conv1 and features of all layers of Trm1 and Trm3 when testing FaST-VGS and FaST-VGS+ on this benchmark.

\section{Results}\label{sec:exp}
\begin{table*}[htb]
  \caption{Comparison of different variants of FaST-VGS and ResDAVEnet~\cite{Hsu2019TransferLF} on the SpokenCOCO 5k test set (the Karpathy split~\citep{Karpathy2017DeepVA}). Subscript CTF means coarse-to-fine retrieval \citep{peng2021fastslow}. SC stands for SpokenCOCO, and LS stands for LibriSpeech.}
  \begin{center}
  \begin{tabular}{lcccccccc}
      \toprule
      &&&\multicolumn{3}{c}{Speech $\rightarrow$ Image}&\multicolumn{3}{c}{Image $\rightarrow$ Speech}\\\cmidrule(lr){4-6} \cmidrule(lr){7-9}
       Model&Pretrained&Data&R@1&R@5&R@10 &R@1&R@5&R@10\\
      \midrule
      ResDAVEnet& &SC &17.3 &41.9 &55.0 & 22.0&50.6 &65.2 \\
      $\text{FaST-VGS}_{\text{CTF}}$& &SC & 30.9&59.7&71.7&42.3&72.2&83.0 \\
      $\text{FaST-VGS+}_{\text{CTF}}$& &LS+SC & 33.1&62.4&74.2&45.3&74.3&84.4 \\
      $\text{FaST-VGS}_{\text{CTF}}$&\checkmark& SC & 35.9&66.3&77.9&48.8&78.2&87.0 \\
      $\text{FaST-VGS+}_{\text{CTF}}$&\checkmark& SC & 35.9&66.3&78.2&48.4&77.0&86.2 \\
      $\text{FaST-VGS+}_{\text{CTF}}$&\checkmark& LS+SC & 35.8&66.5&77.7&48.7&77.8&86.7 \\
      \bottomrule
  \end{tabular}
  \end{center}
  \label{tab:retrieval}
  \end{table*}
 
 \begin{figure}
     \centering
     \includegraphics[width=0.4\textwidth]{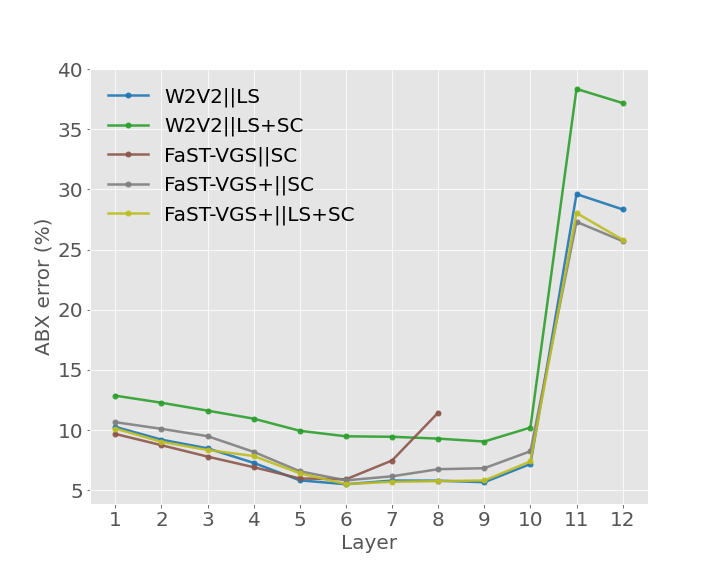}
     \caption{Models on the phonetic task of ZeroSpeech2021 (dev set)}
     \label{fig:pho}
 \end{figure}

\begin{figure*}
     \centering
     \includegraphics[width=0.8\textwidth]{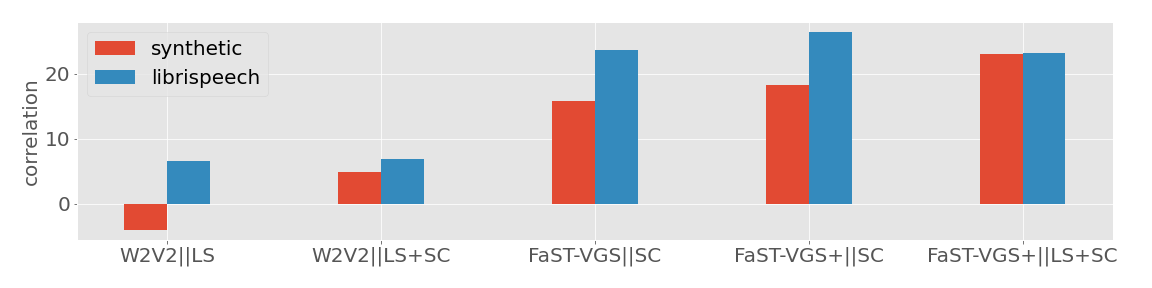}
     \caption{Models on the semantic task of ZeroSpeech2021 (dev set)}
     \label{fig:sem}
 \end{figure*}

\begin{table}[htb]
  \caption{Comparison of different models performance on the lexical and syntactic tasks (dev set) to prob for the impact of different datasets and architectures. SC stands for SpokenCOCO, and LS stands for LibriSpeech. In both cases, higher scores are better (indicated by the upward arrows).}
  \begin{center}
  \resizebox{0.47\textwidth}{!}{
  \begin{tabular}{llccc}
      \toprule
      &&\multicolumn{2}{c}{Lexical $\uparrow$}&\multicolumn{1}{c}{Syntactic $\uparrow$}\\\cmidrule(lr){3-4} \cmidrule(lr){5-5}
       Model&Data&all&in vocab.&\\
      \midrule
      FaST-VGS& SC &63.70&68.95&53.54\\
      FaST-VGS+& SC &65.59&72.52&57.23\\
      FaST-VGS+& LS+SC &67.60&75.43&56.67\\
      \bottomrule
  \end{tabular}}
  \end{center}
  \label{tab:lex_syn_dev}
  \end{table}
\begin{table*}[htb]
  \caption{Performance on the ZeroSpeech 2021 benchmark (test set). Results are shown in percentage except for the semantic task. In the Data column, SC stands for SpokenCOCO, LS stands for LibriSpeech, and PS stands for Places Audio Captions\citep{Harwath2015DeepMS}. H.b. stands for high budget and l.b. stands for low budget. Note that the results in the upper part of the table represent non-VG models, and the lower part of the table represents VG models. VG and non-VG conditions occupy different tracks of the ZeroSpeech competition. For the ABX task, lower scores are better (indicated by the downward arrow), and for the lexical and semantic tasks higher scores are better. For the semantic task, scores further from 0 are better.}
  \begin{center}
  \resizebox{1.0\textwidth}{!}{
  \begin{tabular}{llccccccccc}
      \toprule
&&\multicolumn{4}{c}{Phonetic ABX $\downarrow$}&\multicolumn{2}{c}{Lexical $\uparrow$}&\multicolumn{1}{c}{Syntactic $\uparrow$}&\multicolumn{2}{c}{Semantic (Further from 0)}\\
\cmidrule(lr){3-6}\cmidrule(lr){7-8}\cmidrule(lr){9-9}\cmidrule(lr){10-11}
Model&Data&W-C&W-O&A-C&A-O&all&in vocab.&&Syn.&Lib.\\
      \midrule
      LSTM baseline&LS&3.43&4.81&4.31&7.92&60.55&66.22&52.89&7.35&2.38 \\
      BERT baseline&LS&3.43&4.84& 4.17&7.59&67.66&75.55&56.13&6.25&4.35\\
     \cite{Niekerk21}&LS&5.41&8.67 &6.89&13.14&65.06&72.86&53.95&9.23&-1.14 \\
     \cite{Maekaku2021SpeechRL}&LS&3.15&5.13&4.25&8.64&61.15&66.36&53.88&7.00&-1.47
     \\
     \cite{Chorowski2021InformationRF}&LS&2.85&4.44&3.69&7.28&64.15&72.47&52.55&5.15&-0.85 \\
      \midrule
      VG baseline (l.b.)&LS+SC&8.39&10.66&10.59&15.03&52.86&54.93&53.02&9.71&0.16 \\
      VG baseline (h.b.) &LS+SC&5.36&7.35&6.71&11.92&67.20&74.85&54.53&9.99&-0.10 \\
      Kim et al. &LS+PS&6.50&9.95&9.04&15.44&51.36&51.91&50.43&8.23&16.76\\
      FaST-VGS+ &LS+SC&4.24&5.22&5.08&7.91&67.56&75.23&57.40&15.10&14.32 \\
      \bottomrule
  \end{tabular}}
  \end{center}
  \label{tab:zs21_test}
  \end{table*}
  
\begin{table*}[ht]
\centering
\caption{
Universal speech representation evaluation on SUPERB benchmark. ParaL denotes Paralinguistics aspect of speech. SC stands for SpokenCOCO, LS stands for LibriSpeech, LL stands for Libri-light~\citep{libri-light}, Mix stands for a mixture of speech data collected in~\cite{chen2021wavlm}}\label{tab:superb}
\resizebox{1.0\textwidth}{!}{
\begin{tabular}{lccrrrrrrrrrrrc}

\toprule
\multirow{3}{*}{Method} & \multirow{3}{*}{\#Params} & \multirow{3}{*}{Data}
 & \multicolumn{3}{c}{Speaker} & \multicolumn{5}{c}{Content} & \multicolumn{3}{c}{Semantics}  & ParaL \\ \cmidrule(lr){4-6} \cmidrule(lr){7-11} \cmidrule(lr){12-14} \cmidrule(lr){15-15}

& & & \multicolumn{1}{c}{SID} & \multicolumn{1}{c}{ASV} & \multicolumn{1}{c}{SD} & \multicolumn{1}{c}{PR} & \multicolumn{2}{c}{ASR (WER)} & \multicolumn{1}{c}{KS} & \multicolumn{1}{c}{QbE} & \multicolumn{1}{c}{IC} & \multicolumn{2}{c}{SF} & \multicolumn{1}{c}{ER} \\ \cmidrule(lr){4-15}

& & & \multicolumn{1}{c}{Acc $\uparrow$} & \multicolumn{1}{c}{EER $\downarrow$} & \multicolumn{1}{c}{DER $\downarrow$} & \multicolumn{1}{c}{PER $\downarrow$} & \multicolumn{1}{c}{w/o $\downarrow$} & \multicolumn{1}{c}{w/LM $\downarrow$} &\multicolumn{1}{c}{Acc $\uparrow$} & \multicolumn{1}{c}{MTWV $\uparrow$}  & \multicolumn{1}{c}{Acc $\uparrow$} & \multicolumn{1}{c}{F1 $\uparrow$ }& \multicolumn{1}{c}{CER $\downarrow$} & \multicolumn{1}{c}{Acc $\uparrow$}\\

\midrule
FBANK & 0 & - & 8.5E-4 & 9.56 & 10.05 & 82.01 & 23.18  &15.21& 8.63 & 0.0058 & 9.10 & 69.64 & 52.94 & 35.39  \\ \midrule

PASE+ & 7.83M & LS50& 37.99 & 11.61 & 8.68 & 58.87 & 25.11 &16.62& 82.54 & 0.0072 & 29.82 & 62.14 & 60.17 & 57.86   \\ \midrule

APC & 4.11M & LS360& 60.42 & 8.56 & 10.53 & 41.98 & 21.28 &14.74 & 91.01 & 0.0310 & 74.69 & 70.46 & 50.89 & 59.33   \\

VQ-APC & 4.63M & LS360& 60.15 & 8.72 & 10.45 & 41.08 & 21.20 & 15.21& 91.11 & 0.0251 & 74.48 & 68.53 & 52.91 & 59.66  \\

NPC & 19.38M & LS360& 55.92 & 9.40 & 9.34 & 43.81 & 20.20 & 13.91 & 88.96 & 0.0246 & 69.44 & 72.79 & 48.44 & 59.08  \\

Mockingjay & 85.12M & LS360& 32.29 & 11.66 & 10.54 & 70.19 & 22.82 & 15.48  & 83.67 & 6.6E-04 & 34.33 & 61.59 & 58.89 & 50.28  \\

TERA & 21.33M & LS360& 57.57 & 15.89 & 9.96 & 49.17 & 18.17 & 12.16 & 89.48 & 0.0013 & 58.42 & 67.50 & 54.17 & 56.27   \\

wav2vec & 32.54M & LS960& 56.56 & 7.99 & 9.9 & 31.58 & 15.86 & 11.00 & 95.59 & 0.0485 & 84.92 & 76.37 & 43.71 & 59.79  \\

vq-wav2vec & 34.15M & LS960& 38.80 & 10.38 & 9.93 & 33.48 & 17.71 & 12.80 & 93.38 & 0.0410 & 85.68 & 77.68 & 41.54 & 58.24  \\

wav2vec 2.0 Base & 95.04M & LS960& 75.18  & 6.02 & 6.08 & 5.74 & 6.43 & 4.79 & 96.23 & 0.0233 & 92.35 & 88.30 & 24.77 & 63.43  \\

HuBERT Base & 94.68M & LS960& 81.42 & 5.11 & 5.88  & 5.41 & 6.42 & 4.97 & 96.30 & 0.0736 & 98.34 & 88.53 & 25.20 & 64.92  \\

\midrule
FaST-VGS & 187.87M & LS960+SC742 & 41.49&6.54&6.50&16.30&13.46&9.51&96.85&0.0546&98.37&84.91&32.33&57.37 \\
FaST-VGS+ & 217.23M & LS960+SC742 & 41.34&5.87&6.05&7.76&8.83&6.37&97.27&0.0562&98.97&88.15&27.12&60.96  \\
\midrule

modified CPC & 1.84M & LL60k& 39.63 & 12.86 & 10.38 & 42.54 & 20.18 &13.53 & 91.88 & 0.0326 & 64.09 & 71.19 & 49.91 & 60.96  \\
WavLM Base+ & 94.70M & Mix94k&86.84&	4.26&	4.07&	4.07&	5.64& ---- &	96.69&	\textbf{0.0990}	&\textbf{99.16}&	89.73	&21.54&	67.98 \\

wav2vec 2.0 Large & 317.38M & LL60k& 86.14  & 5.65 &  5.62 & 4.75 & 3.75 & 3.10 & 96.66 & 0.0489 & 95.28 & 87.11 & 27.31 & 65.64   \\

HuBERT Large & 316.61M & LL60k& 90.33 & 5.98 & 5.75 & 3.53 & 3.62 & 2.94 & 95.29 & 0.0353 &98.76 & 89.81 & 21.76 & 67.62  \\

WavLM Large & 316.62M & Mix94k& \textbf{95.25}	& \textbf{4.04} &	\textbf{3.47}	& \textbf{3.09}& \textbf{3.51}	& ---- &	\textbf{97.40}&	0.0827&	99.10 &	\textbf{92.25}& \textbf{17.61}&	 \textbf{70.03} 
 \\

\bottomrule
\end{tabular}}

\end{table*}

\raggedbottom

\subsection{Speech-Image Retrieval}
Tab.~\ref{tab:retrieval} shows speech-image retrieval results on the test set of SpokenCOCO. We use CTF retrieval for FaST-VGS and FaST-VGS+. Comparing the second and third row, we see that when we do not load the conv1, trm1, and trm3 layers weights from pretrained wav2vec2.0 and trained the all components from scratch (except for the image feature extractor Faster-RCNN), FaST-VGS+ outperforms FaST-VGS by a large margin. This shows the benefit of the addition of the masked language modeling objective on the speech-image retrieval task.
When we load pretrained weights from wav2vec2.0, with FaST-VGS+, we see small improvements in speech to image retrieval and small degredations in image-to-speech retrieval as compared to FaST-VGS. However, in both cases these differences are very slight. Comparing these models to those trained without wav2vec2.0 initialization, we see a large difference in performance (i.e. FaST-VGS+ with pretrained weights achieves average R@10 82.2, FaST-VGS+ trained from scratch achieves average R@10 79.3). We believe that this is due to the fact that the wav2vec2.0 Base model underwent many more total epochs of training than our FaST-VGS+ models trained without wav2vec2.0 initialization, and therefore already converged to a very good optimum; this would also explain why we did not see significant differences in retrieval performance between the FaST-VGS and FaST-VGS+ models that were both initialized with wav2vec2.0 weights.

\subsection{ZeroSpeech 2021}
\textbf{Phonetic}. In Fig.\ref{fig:pho}, we show the average ABX error of representations from different layers of different models on four splits of the dev set. Wav2vec2.0 trained on LibriSpeech performs very good on this task, however, when we further trained the model on SpokenCOCO, we see a significant increase in the ABX error. Since the phonetic task evaluation data is sampled from LibriSpeech, the increase in ABX error reflects the impact of domain shift. On the other hand, FaST-VGS and FaST-VGS+ trained on SpokenCOCO only achieves comparable performance with wav2vec2.0 trained on LibriSpeech. FaST-VGS+ trained on both SpokenCOCO and LibriSpeech achieves similar performance to wav2vec2.0, except for the last two layers, where we see slightly better performance from FaST-VGS+. Given the fact that FaST-VGS and FaST-VGS+ are initialized from wav2vec2.0 trained on LibriSpeech, these results demonstrate and that visually grounding can reduce the negative impact of domain shift on this task as compared to an audio-only model. 

\textbf{Semantic}. Using the same model configurations we used for the ABX task, we visualize their performance on the ZeroSpeech semantic task in Fig.~\ref{fig:sem}. This task requires the model's output to be pooled temporally, and we consider both mean and max pooling. We also consider features extracted from all layers in Trm1, Conv2 and Trm2. Only the best configuration for each model is shown in Fig.~\ref{fig:sem}. We found that for wav2vec2.0 trained on LibriSpeech, mean pooling on features from the first layer of Trm1 performs the best; for wav2vec2.0 trained on LibriSpeech and the audio of SpokenCOC0, mean pooling on features from layer 8 of Trm1 performs the best; for FaST-VGS, mean pooling on features of layer 8 of Trm1 performs the best; for FaST-VGS+ (either trained on SpokeCOCO and LibriSpeech or on LibriSpeech only), max pooling on features from layer 1 of Trm2 performs the best. We see that in general, the VG models significantly outperform the non-VG models on this task. For the two FaST-VGS+ results, since all correlations are above $0$, we can take the average to compare them, and FaST-VGS+ trained on both SpokenCOCO and LibriSpeech slightly outperform FaST-VGS+ trained on LibriSpeech only ($23.09$ v.s. $22.32$). 


\textbf{Lexical and syntactic}. As shown in Tab.~\ref{tab:lex_syn_dev}, FaST-VGS+ significantly outperforms FaST-VGS on the lexical and syntactic tasks, which shows that combining cross-modal contrastive loss and unimodal audio constrastive loss is better than using cross-modal contrastive loss only. Similar to the phonetic and semantic tasks, the difference of FaST-VGS+ trained on only SpokenCOCO and on both SpokenCOCO and LibriSpeech is small, indicating that the dataset has a smaller effect on the model performance.

\textbf{Comparisons to ZeroSpeech leaderboard.} We compare the performance of FaST-VGS+ on all of the ZeroSpeech tasks against several baselines and top-performing submissions in Tab.~\ref{tab:zs21_test}\footnote{as of 11/19/2021, on https://zerospeech.com/2021/results.html}. The upper part lists leading submissions of non-VG models and lower part are VG models. Note that in the table, only BERT baseline, VG baseline (h.b.), and our FaST-VGS+ are submissions to the high budget track, and all other models are in the low budget track. For the phonetic task, we see that non-VG models generally performs better than VG models, but our FaST-VGS+ ranks the first among VG models; it should also be noted that per the official ZeroSpeech task description, ABX scores of 5 and below are considered to indicate strong performance. For the lexical task, FaST-VGS+ ranks first among VG models and is only slightly worse than the BERT Baseline. For the syntactic task, FaST-VGS outperforms all models, which shows the benefit of incorporating visual context in unsupervised language modeling. For the semantic task, in general the VG models outperform the non-VG models, but FaST-VGS+ performs consistently well between both synthetic and librispeech data, and beats Kim et al. in terms of the average score ($14.71$ v.s. $12.50$).


\subsection{SUPERB}
Our results on the SUPERB benchmark are shown in Tab.~\ref{tab:superb}. The models are ordered by the amount of audio data used during training. We see that the two VG models achieve poor performance in speaker identification (SID), and we hypothesize that this is partly due to the fact that SpokenCOCO contains multiple speakers describing the same image, and since the similarity of audio from different speaker and the same image should be similar, speaker information is filtered out. Compared to their initialization, i.e. wav2vec2.0 base, the two VG models also perform worse on other speaker modeling tasks such as automatic speaker verification (ASV) and speaker diarization (SD). For content modeling, the two VG models perform worse than wav2vec 2.0 Base and HuBERT Base, and we hypothesize that this is because the multimodal grounding objective encourages the model to emphasize higher level linguistic information (such as words, phrases, and semantics), and therefore character recognition based on CTC may suffer. We will leave the investigation of this issue for future work. FaST-VGS+ outperforms other models that are trained solely on LS960, and ranks the second overall in keyword spotting (KS), which shows that visual grounding can drastically reduce the need for audio data (1.7k hours v.s. 94k hours) for this task. The strong performance on KS also supports our hypothesis that word level information is effectively captured by the learned representations. Our VG models also perform well on intent classification (IC), where FaST-VGS+ ranks third overall. 

In general we see the benefit of incorporating visual grounding in some of the tasks (e.g. KS and IC), but whether visual information can reduce the need to the amount of audio data for other tasks requires further investigation. 

\section{Conclusion}
In this paper, we explored the benefit of incorporating visual context for learning speech representations using two VGS models, FaST-VGS~\cite{peng2021fastslow} and the novel FaST-VGS+. We evaluated our approach on the ZeroSpeech 2021 and SUPERB benchmarks, and show that for some tasks, visual grounding can offer benefits over audio-only models. We also find that FaST-VGS+ outperforms FaST-VGS in many tasks, indicating the potential benefits of combining masked language modeling with visual grounding. Our results suggest that visual grounding can help the learning of higher level, semantic structure from the speech signal, as evidenced by the performance of FaST-VGS+ on the ZeroSpeech semantic task and the keyword spotting and intent classification SUPERB tasks. However, surprisingly and contrary to our intuition, our results also suggest that this ability does not seem to greatly benefit tasks that focus on sub-word structure, such as the ZeroSpeech ABX task or the phone recognition and ASR tasks in SUPERB. In our future work, we plan to investigate improved strategies for combining visual grounding and masked language modeling objectives within self-supervised speech models.

\bibliography{aaai22}

\end{document}